\documentstyle[12pt]{article}
\date{}
\begin{document}

\begin{center}{\Large\bf A lower estimate
for the modified Steiner functional}
\end{center}
\vskip 2cm
\begin{center}{\large George K. Savvidy}
\end{center}
\begin{center}Physics Department,University of Crete,71409 Iraklion,Greece
\end{center}
\begin{center}Institut f\"ur Theoretische Physik,D-60325 Frankfurt,Germany
\end{center}
\begin{center}e-mail: savvidy@knosos.cc.uch.gr
\end{center}
\begin{center}and
\end{center}
\begin{center}{\large Rolf Schneider}
\end{center}
\begin{center}Mathimatisches Institut, Universitat Freiburg,
Albertstr.23b, 7800 Freiburg, Germany
\end{center}
\begin{center}e-mail: rschnei@sun1.ruf.uni-freiburg.de
\end{center}
\sloppy
\addtolength{\topmargin}{-60pt}
\addtolength{\textheight}{150pt}
\textwidth165mm
\addtolength{\oddsidemargin}{-17mm}
\jot3mm

\vskip 5cm
\begin{center}{\large\bf Absrtact}
\end{center}

  We prove inequality (1) for the modified Steiner functional A(M),
which extends the notion of the integral of mean curvature for
convex surfaces. We also establish an expression for A(M)
in terms of an integral over all hyperplanes intersecting the polyhedral
surface M.

\newpage
{\bf 1. Introduction}
\vspace{.5cm}

In the articles \cite{Amb}, \cite{Sav1}, \cite{Sav2} the authors suggest a
new version of string theory,
which  can  be
considered as a natural extension of the Feynman--Kac integral over
paths to an integral over
surfaces.
Both amplitudes coinside in the case, when the surface degenarates
into a single partical world line.

The  string  has been conjectured to describe a wide variety
of physical phenomena,
including strong interaction, the three dimensional Ising model,
and unified models incorporating gravity.
The Feynman integral for the string is just the partition function
for the randomly
fluctuating surfaces, and in this
statistical approach the surface is associated with
a connected polyhedral
surface embedded in euclidean space.

To prove the convergence of the partition function
for this new string, the authors of
\cite{Sav1}, \cite{Sav2}
require a lower estimate
for the action $A(M)$ on which the theory is based. The purpose of
the present note is to prove
the inequality
\begin{equation} \label{1}
A(M) > 2\pi \Delta,
\end{equation}
where $A(M)$ is the modified Steiner functional as introduced in \cite{Amb},
\cite{Sav1}, \cite{Sav2} and
$\Delta$ is the
diameter of the polyhedral surface $M$ in ${\sf R}^{d}$.
We also establish an expression for A(M) in
terms of an integral over all hyperplanes intersecting the
polyhedral surface M.
\vspace{1cm}

{\bf 2.Proof of the inequality}
\vspace{.5cm}

We recall the definition of $A(M)$.
\vspace{.5cm}

{\bf Definition.}Let $M$ be an embedded
connected closed polyhedral surface in
euclidean space ${\sf R}^{d}$ $(d \ge 3)$. Let ${\cal F}_{1}(M)$
be the set of edges of $M$. For
$e \in {\cal F}_{1}(M)$ we denote by $L(e)$ the length of $e$
and by $\alpha(e)$, where $0 <
\alpha(e) < \pi$, the angle between the two faces of $M$ incident with $e$.
Then the modified
Steiner functional is defined by
\begin{displaymath}
A(M) := \sum_{e \in {\cal F}_{1}(M)} L(e)[\pi - \alpha(e)].
\end{displaymath}
\vspace{.5cm}

{\bf Theorem.} {\it If $\Delta$ denotes the diameter of M, then}
\begin{equation} A(M) > 2\pi \Delta.
\end{equation}

{\it Proof.} We first consider a simple closed polygon $P$ in ${\sf R}^{d}$.
For a
vertex $v$ of $P$,
we denote by $\alpha(v)$, where $0 < \alpha(v) < \pi$,
the angle between the
two edges of $P$ incident with $v$.
The (absolute) total curvature of $P$ is defined by
\begin{displaymath}
\kappa(P) := \sum_{v}[\pi - \alpha(v)],
\end{displaymath}
where the sum extends over the vertices of $P$. It is known that
\begin{equation}\label{2}
\kappa(P) \ge 2\pi
\end{equation}
(Fenchel's inequality for polygons; see, e.g., \cite{Mil}).

In the proof of inequality (\ref{1}) we
shall use some integral geometry, in particular the space
${\cal E}^{d}_{d-1}$ of hyperplanes in
${\sf R}^{d}$ with its (suitably normalized) rigid motion
invariant measure $\mu_{d-1}$; see, e.g., \cite{Sch}.
According to \cite{Sch}, (1.9), the measure
$\mu_{d-1}$
can be represented as follows.
For a nonnegative measurable function $f$ on ${\cal E}^{d}_{d-1}$
we have
\begin{displaymath}
\int \limits_{{\cal E}^{d}_{d-1}}fd\mu_{d-1} =
\int \limits_{S^{d-1}}\int \limits_{-\infty}
^{\infty}f(H_{u,\tau}) d\tau d\sigma(u).
\end{displaymath}
Here $S^{d-1} := \{u \in {\sf R}^{d}: \|u\| = 1\}$
is the unit sphere of ${\sf R}^{d}$,
\begin{displaymath}
H_{u,\tau} := \{x \in {\sf R}^{d}: \langle x,u \rangle =
\tau\}, \quad u \in S^{d-1}, \enspace
\tau \in {\sf R},
\end{displaymath}
is a general hyperplane with unit normal vector $u$,
and $\sigma$ is the spherical Lebesgue
measure on $S^{d-1}$, normalized to total measure 1.
By $\langle \cdot, \cdot \rangle$ we denote
the scalar product in ${\sf R}^{d}$.

The hyperplane $H \in {\cal E}^{d}_{d-1}$ is said to
intersect the polyhedral surface $M$
{\em in general position} if $H \cap M \not= \emptyset$
and $H$ does not contain a vertex of $M$.
In that case, the intersection of $H$ with
an edge of $M$ is either empty or a point, and the
intersection of $H$ with a face of $M$ is either empty or a segment.
It follows that the
intersection $H \cap M$ is the union of finitely
many simple closed polygons $P_{1}(H),\dots,
P_{k}(H)$, and from inequality (\ref{2})
(applied in $H$ instead of ${\sf R}^{d}$) we have
\begin{displaymath}
\kappa(H \cap M) := \kappa(P_{1}(H)) + \dots + \kappa(P_{k}(H)) \ge 2\pi.
\end{displaymath}
It follows that
\begin{equation}\label{3}
I  := \int \limits_{{\cal E}^{d}_{d-1}} \kappa(H \cap M) d\mu_{d-1}(H)
\nonumber
\ge  2\pi \mu_{d-1}(\{H \in {\cal E}^{d}_{d-1} : H \cap M \not= \emptyset\}),
\end{equation}
since the set of all hyperplanes intersecting $M$, but not in general
position, has $\mu_{d-1}$-
measure zero. Let $S$ be a segment connecting two points
of $M$ with maximal distance, so that
the length of $S$ is equal to the diameter $\Delta$ of $M$.
Let $s$ be a unit vector parallel to
$S$. Then
\begin{eqnarray*}
& & \mu_{d-1}(\{H \in {\cal E}^{d}_{d-1}: H \cap M \not= \emptyset\})
 =  \int \limits_{S^{d-1}} \int
\limits_{-\infty}^{\infty}{\bf 1}_{\{H_{u,\tau}\cap M \not=
\emptyset\}} d\tau d\sigma(u) \\
& & >  \int \limits_{S^{d-1}}
\int \limits_{-\infty}^{\infty}{\bf 1}_{\{H_{u,\tau} \cap S \not=
\emptyset\}} d\tau d\sigma(u)
 =  \int \limits_{S^{d-1}}\Delta | \langle u,s \rangle| d\sigma(u) \\
&  & = c_{1}\Delta.
\end{eqnarray*}
Here ${\bf 1}_X$ denotes the indicator function of $X$.
By $c_{1},\dots,c_{6}$ we denote constants depending only
on the dimension $d$. We have proved
that
\begin{equation}\label{4}
I > c_{2}\Delta.
\end{equation}

On the other hand, if the hyperplane $H$ intersects $M$
in general position, we can write
\begin{equation}\label{5}
\kappa(H \cap M) = \sum_{e \in {\cal F}_{1}(M)} [\pi - \beta(M,e,H)],
\end{equation}
where $\beta(M,e,H)$ is defined as follows.
If $H$ meets the edge $e$ (and hence the relative
interior of $e$) and if $F_{1},F_{2}$ are the two
faces of $M$ incident with $e$, then
$\beta(M,e,H) \in (0,\pi)$ is the angle between the
segments $H \cap F_{1}$ and $H \cap F_{2}$
at the point $H \cap e$. If $H$ does not meet $e$,
we put $\beta(M,e,H) = \pi$. We can now write
\begin{equation}\label{6}
I = \sum_{e \in {\cal F}_{1}(M)} \;
\int \limits_{{\cal E}^{d}_{d-1}}[\pi - \beta(M,e,H)]
d\mu_{d-1}(H).
\end{equation}
Let $e \in {\cal F}_{1}(M)$ be a fixed edge. We have
\begin{eqnarray}\label{7}
&   & \int \limits_{{\cal E}^{d}_{d-1}}
[\pi - \beta(M,e,H)]d\mu_{d-1}(H) \nonumber \\
& = & \int \limits_{S^{d-1}}
\int \limits_{-\infty}^{\infty}[\pi - \beta(M,e,H_{u,\tau})]
d\tau d\sigma(u)  \nonumber \\
& = & \int \limits_{S^{d-1}}
\left[\pi - \beta(M,e,H_{u,\langle x,u \rangle}\right]L(e) |
\langle u,w(e) \rangle| d\sigma(u),
\end{eqnarray}
where $x$ is some fixed point in the relative interior
of the edge $e$ and $w(e)$ denotes a unit
vector parallel to the edge $e$.

Let $F_{1}, F_{2}$ be the two faces of $M$ incident with
$e$ and let $\alpha(e) \in (0,\pi)$ be
the angle between $F_{1}$ and $F_{2}$, as defined initially. We assert that
\begin{equation} \label{8}
\int \limits_{S^{d-1}}\left[\pi - \beta(M,e,H_{u,\langle x,u \rangle})\right]
 | \langle u,w(e)
\rangle | d \sigma(u) = c_{5}[\pi - \alpha(e)].
\end{equation}
For the proof we may assume, without loss of generality,
that $x$ is the origin of ${\sf
R}^{d}$. The integral in (\ref{8}) can be written in the form
\begin{equation} \label{9}
J := \int \limits_{{\cal L}^{d}_{d-1}}f(H)d\nu_{d-1}(H),
\end{equation}
where ${\cal L}^{d}_{d-1}$ denotes the space of
$(d-1)$-dimensional linear subspaces of ${\sf R}
^{d}$ and $\nu_{d-1}$ is its normalized invariant measure;
the function $f$ is defined by
\begin{displaymath}
f(H) = [\pi - \beta(M,e,H)] | \langle u_{H},w(e) \rangle |,
\end{displaymath}
where $u_{H}$ is a unit normal vector of $H$.
The edge $e$ and the two adjacent faces $F_{1},
F_{2}$ of $M$ lie in a 3-dimensional linear subspace $A$ of ${\sf R}^{d}$.
For $\nu_{d-1}$-
almost all $H \in {\cal L}^{d}_{d-1}$,
the intersection $A \cap H$ is a 2-dimensional linear
subspace. In that case, the angle $\beta(M,e,H)$ depends
only on $M$ and this subspace, so
that we can write $\beta(M,e,H) = \beta(M,e,A \cap H)$. Moreover,
\begin{displaymath}
|\langle u_{H},w(e) \rangle | =
| \langle u_{H},u_{A \cap H} \rangle \langle u_{A \cap H},
w(e) \rangle |,
\end{displaymath}
where $u_{A \cap H}$ is a unit normal vector of $A \cap H$ in $A$.

Using a general formula of integral geometry, one can write the
integral (\ref{9}) in the form
\[ \int \limits_{{\cal L}^{d}_{d-1}}f(H)d\nu_{d-1}(H)
 =  c_{3} \int \limits_{{\cal L}^{A}_{2}}
\int \limits_{{\cal L}^{L}_{d-1}}f(H)[H,A]^{2}
d\nu_{d-1}^{L}(H)d\nu_{2}^{A}(L).
\]
Here ${\cal L}^{A}_{2}$ denotes the space of 2-dimensional linear
subspaces of $A$  and $\nu_{2}
^{A}$ is the normalized invariant measure on this space.
For fixed $L \in {\cal L}^{A}_{2}$,
${\cal L}^{L}_{d-1}$ denotes the space of hyperplanes
containing $L$, and $\nu_{d-1}^{L}$ is
the invariant measure on this space. $[H,A]$
is a certain function depending only on the relative
position of $H$ and $A$; it is invariant under simultaneous
rotations of $H$ and $A$. The
identity above is equivalent to a special case of formula (14.40)
in Santal\'o \cite{San}, but written
in the style of \cite{Sch}. Applying this to our present situation, we obtain
\begin{eqnarray*}
J & = & c_{3} \int \limits_{{\cal L}^{A}_{2}}
\int \limits_{{\cal L}^{L}_{d-1}}[\pi - \beta
(M,e,H)] \,| \langle u_{H}, u_{A \cap H}\rangle
\langle u_{A \cap H},w(e) \rangle | \,
[H,A]^{2}d\nu_{d-1}^{L}(H)d\nu_{2}^{A}(L) \\
& = & c_{3} \int \limits_{{\cal L}^{A}_{2}}
[\pi - \beta(M,e,L)] \,| \langle u_{L},w(e) \rangle |
\int \limits_{{\cal L}^{L}_{d-1}} [H,A]^{2} \,| \langle u_{H},u_{L}\rangle | \,
d \nu_{d-1}^{L}(H)d\nu_{2}^{A}(L) \\
& = & c_{4} \int \limits_{{\cal L}^{A}_{2}}[\pi - \beta(M,e,L)] \,
| \langle u_{L},w(e) \rangle | \,
d\nu_{2}^{A}(L).
\end{eqnarray*}
The final integral is a mean value over 2-dimensional linear
subspaces in a 3-dimensional
euclidean space. Its value can be obtained from the more general
Theorem 3.2.1 in \cite{Sch}. In this
way we arrive at
\begin{displaymath}
J = c_{5}[\pi - \alpha(e)],
\end{displaymath}
which proves (\ref{8}).

Taking (\ref{3}), (\ref{6}), (\ref{7}), (\ref{8}) together, we deduce that
\begin{equation} \label{10}
A(M) > c_{6}\Delta.
\end{equation}
In order to find the optimal constant $c_{6}$ for
which (\ref{10}) holds generally, we consider the
boundary $M_{\epsilon}$ of a triangular prism with
height $\Delta$ and base a regular triangle with
edge length $\epsilon$. For $\epsilon \rightarrow 0$,
the diameter of $M_{\epsilon}$ tends to
$\Delta$ and $A(M_{\epsilon})$ tends to $2\pi \Delta$.
It follows that $c_{6} \le 2\pi$. On the
other hand, from the way inequality (\ref{10})
was obtained it is easy to see that $A(M) >
A(M_{\epsilon})$, if $\epsilon > 0$ is sufficiently small.
Thus $c_{6} = 2\pi$ is the optimal
constant.
\vspace{1cm}

{\bf 3. Concluding Remark}
\vspace{.5cm}

We want to stress that the represantaton (7) is very useful
for studing more complicated models \cite {Amb}, \cite {Sav1},
\cite {Sav2} and various phenomena.

This wark was supported in part by Alexander von Humboldt Foundation.

\end{document}